\documentclass[aps,twocolumn,superscriptaddress]{revtex4-1}

\usepackage{amsmath,amsfonts,amssymb,amsthm,epsfig,array,physics}
\usepackage{graphicx,tabularx}
\usepackage{dsfont}
\usepackage{upgreek}
\usepackage[etex=true,export]{adjustbox}
\usepackage{graphics}
\usepackage{float}
\usepackage{bm}
\usepackage{verbatim}
\usepackage{gensymb}
\usepackage[dvipsnames]{xcolor}
\usepackage{slashed}
\usepackage[hidelinks,colorlinks,linkcolor=blue,
citecolor=blue,urlcolor=blue]{hyperref}
\usepackage[titletoc,title]{appendix}

\newcommand{\beq}{\begin{equation}}
\newcommand{\eeq}{\end{equation}}
\newcommand{\beql}{\begin{equation*}}
\newcommand{\eeql}{\end{equation*}}
\newcommand{\beqn}{\begin{eqnarray}}
\newcommand{\eeqn}{\end{eqnarray}}

\begin{document}


\title{Electrostatic effects of the MnBi$_2$Te$_4$-superconductor hetero-structures in chiral Majorana search}

\author{Li Chen}
\affiliation{State Key Laboratory of Low Dimensional Quantum Physics, Department of Physics, Tsinghua University, Beijing, 100084, China}
\affiliation{Frontier Science Center for Quantum Information, Beijing 100184, China}

\author{Zhan Cao}
\affiliation{Beijing Academy of Quantum Information Sciences, Beijing 100193, China}


\author{Ke He}
\affiliation{State Key Laboratory of Low Dimensional Quantum Physics, Department of Physics, Tsinghua University, Beijing, 100084, China}
\affiliation{Frontier Science Center for Quantum Information, Beijing 100184, China}
\affiliation{Beijing Academy of Quantum Information Sciences, Beijing 100193, China}

\author{Xin Liu}
\email{phyliuxin@hust.edu.cn}
\affiliation{School of Physics, Huazhong University of Science and Technology, Wuhan, Hubei 430074, China}

\author{Dong E. Liu}
\email{dongeliu@mail.tsinghua.edu.cn}
\affiliation{State Key Laboratory of Low Dimensional Quantum Physics, Department of Physics, Tsinghua University, Beijing, 100084, China}
\affiliation{Frontier Science Center for Quantum Information, Beijing 100184, China}
\affiliation{Beijing Academy of Quantum Information Sciences, Beijing 100193, China}

\date{\today}

\begin{abstract}
    The realization of chiral Majorana modes using hetero-structures is a challenging task. A significant reason is that the previous theoretical models are simple and cannot capture the real physics among the interplay of superconductivity, magnetism, and the electrostatic environment. Beyond the well-known minimal models, we develop a self-consistent Schr\"odinger-Poisson to include a key focus---the electrostatic effects induced by the gate control. We show that electrostatic environment imposes constraints on both induced superconductivity and the effective magnetization, and therefore significantly changes the topological region compared to previous work. However, within our theory, we identify the topological regimes supporting the chiral Majorana mode with practical tunability. Importantly, the induced superconductivity in the topological regime, contrary to traditional beliefs, will not be reduced by the presence of the magnetization. Our results deeply comprehend the real phase diagrams and parameter tunability of the actual devices in chiral Majorana search. 
    
    

\end{abstract}

\maketitle

\section{Introduction}
The chiral Majorana modes (CMMs)~\cite{Read-2000-PRB,Schnyder-2008,Qi-PRL2009,Qi2011} can be considered as the one-dimensional homologous counterpart of Majorana zero modes (MZMs)~\cite{Read-2000-PRB,Kitaev2001}, and are potentially useful for quantum information processing~\cite{Nayak-RevModPhys-2008,Lian-PNAS-2018,Beenakker-PRL-2019}. Pioneering theoretical proposals~\cite{Qi2011,Wangjing2015,WangJi2016} predict that CMMs can be realized in hybrid systems that combine quantum anomalous Hall insulators (QAHI) (please refer to the theories~\cite{Haldane1988,Nagaosa2003The,Qi-Wu-Zhang-2006,Qi-Hughes-Zhang-PRB2008,Liu2008,yu2010} and experiments~\cite{Chang2013}) with superconductors. The half-quantized conductance plateau was proposed to be an evidence for CMMs~\cite{Chung-PRB2011,Wangjing2015,He2017}.
However, a controversy arises because certain non-Majorana trivial mechanisms can also generate similar signatures, especially in disordered samples\cite{Huang2018,Ji2018,Kayyalha2020}. 
In contrast to magnetically doped topological
insulators, the recent discovered  MnBi$_2$Te$_4$ (MBT) family of materials promises a bigger magnetic exchange
gap and less disorders~\cite{Li2019s,Zhang2019PRL,Gong_2019,Yan2019,Otrokov-nature-2019,Chen-NC-2019,Chen-PRX-2019,Li-PRX-2019,Lee-PRR-2019,Vidal_2019,Deng2020,Shikin-Sci_report-2020,Nevola-PRL-2020,Liu2020,Nevola2020,Liu2021NC,Ge-PRL-2022}, which is proposed as an potential platforms to realize CMMs~\cite{Peng2019PRB,Zhang-PRB-2021}.


Another serious problem is that the proposed systems require the coexistence of superconductivity and magnetism, and we may wonder if the CMMs phase can be realized via a feasible parameter control of the device. Previous theoretical works~\cite{Wangjing2015,WangJi2016,Peng2019PRB,Yan-PRB-2019,He-CP-2019,YZB-PRB-2019,Sun2020,Zhang-PRB-2021,Hogl-PRB-2020,Luo-PRB-2021,Zhang-PRL-2021} only considered the simple minimal models, which regard phenomenological parameters, such as chemical potential and induced superconducting (SC) gap, as independently adjustable parameters. Actually, these crucial parameters are highly correlated and cannot be freely tuned in real experiments by controlling the electrostatic environment~\cite{Vuik-IOP-2016,reeg-prb-2017,reeg-prb-2018,Antipov2018,Mikkelsen2018}. This could greatly narrow the topological region and complicates the experimental implementations. 
Thus to understand the device control capability, we need to develop a more reliable numerical simulation scheme for realistic experimental setups, especially for treating both SC proximity effect and the magnetism.

\begin{figure}[t]
\centerline{\includegraphics[width=1\columnwidth]{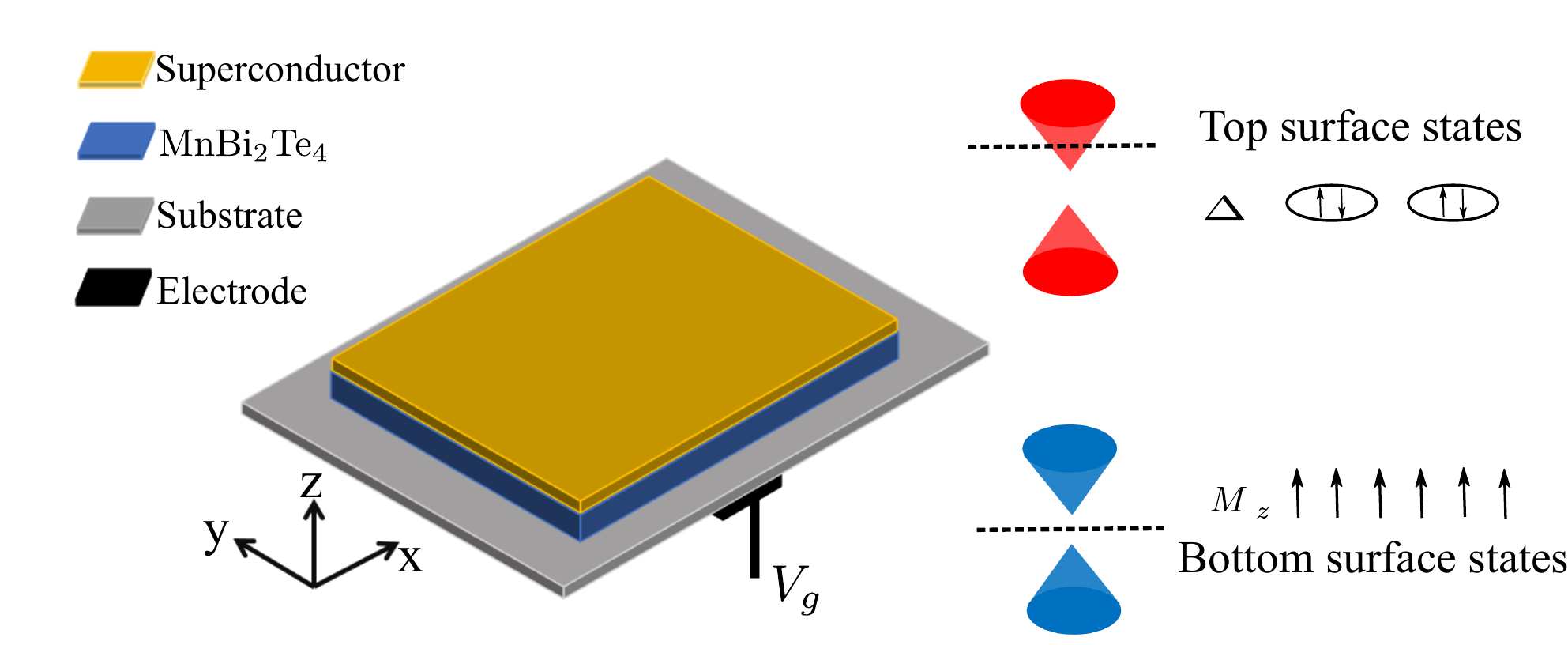}}
\caption{   A MnBi$_2$Te$_4$ thin film is coupled to a s-wave SC on the top surface. The magnetic gap of top surface state is always below the Fermi level during the gate
tuning. CMMs will exist if the Fermi level is tuned in the magnetic gap of bottom surface states. }
 \label{set_up}
\end{figure}

\begin{figure*}[t]
\centerline{\includegraphics[width=2\columnwidth]{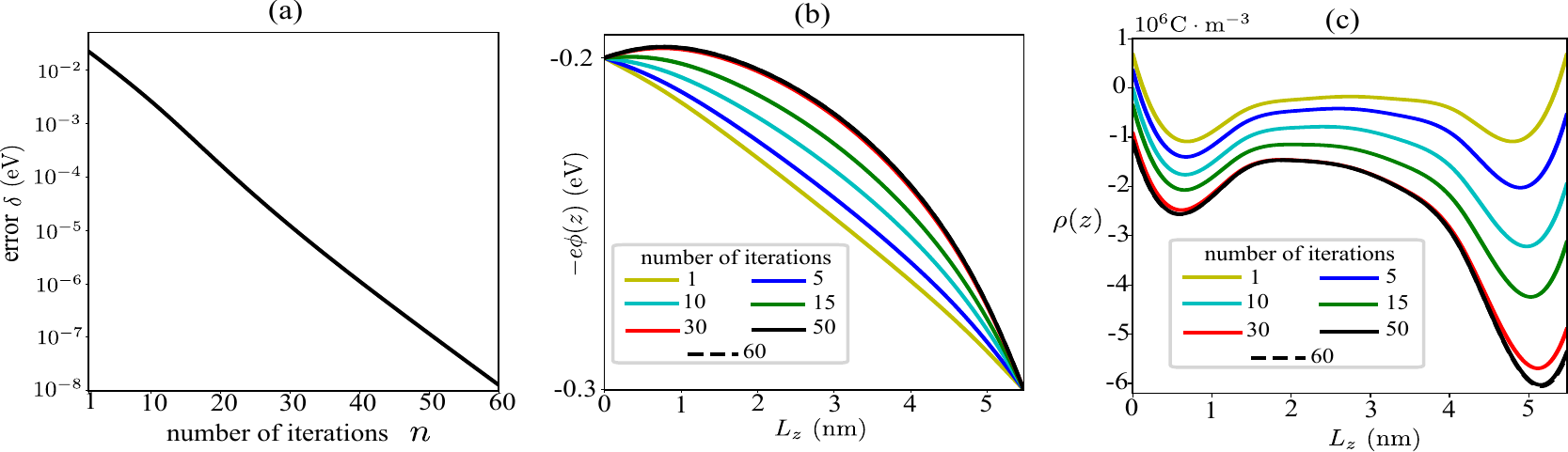}}
\caption{(a) The error of Schr\"odinger-Poisson equations as a function of the number of iterations $n$. The distribution of (b) electrostatic energy $-e\phi(z)$ and (c) carrier density $\rho(z)$ as the the number of iterations increase. The convergence occurs when the iterations number $n>50$ with the error $\sigma < 10^{-7}$ eV, see the black solid and dash lines in panels (b) and (c).}
\label{supp-SP}
\end{figure*}
 
In this work, we consider the debate about CMM realizations and the actual device tunability. Beyond the well-known minimal models, we developed a self-consistent Schr\"odinger-Poisson method \cite{Tan-JAP-1990,Luscombe-PRB-1992,Ambrosetti2008,Vuik-IOP-2016,reeg-prb-2017,reeg-prb-2018,Antipov2018,Mikkelsen2018} to solve the electrostatic problems induced by the actual gate control. We apply this method to study an MBT thin film coupled to an s-wave superconductor (SC) as an example. We find that the band bending effect~\cite{Bardeen-PR-1947,Heine-PR-1965,Gotoh2003,Akhgar2022} at the MBT-SC interface severely restricts the tunability of top surface states, and the corresponding magnetic Dirac gap is always below the Fermi level during the gate tuning. Our results also show that the induced superconductivity varies considerably as tuning the gate voltage. These constraints, which are not properly considered in previous works, are naturally thought to be detrimental to the realization of CMMs.
However, we show that the CMMs can be realized in a reasonable range of experimental parameters. The key point is to tune the Fermi level of the bottom surface state into the magnetic Dirac gap, which is ensured by the high tunability. Remarkably, the required proximity superconductivity will not be reduced by the presence of the magnetization in the topological regime, and ensured a large topological gap. In addition, the previously predicted CMMs phase with Chern number $\mathcal{C}=2$~\cite{Qi2011,Wangjing2015} can not be realized in a real MBT-SC device.

The rest of the paper is organized as follows. In Sec.~\ref{sec_model},  
we construct a model Hamiltonian and calculate the electrostatic potential using Schr\"odinger-Poisson method.  In Sec.~\ref{proximity}, we investigate the proximity effect in MBT-SC hybrid system. In Sec.~\ref{chiral-MBS}, we discuss device control capabilities in the chiral Majorana search, and demonstrate that the key point for achieving CMMs is to tune the chemical potential of BSSs in their magnetic gap.  Finally, we conclude in Sec.~\ref{Conclusion}.

\section{Model Hamiltonian and electrostatic potential}\label{sec_model}
We consider a two dimensional (2D) MBT thin film coupled to an s-wave SC, as shown in Fig.~\ref{set_up}. The antiferromagnetic ordering and the magnetization direction
are both assumed to be along the $z$ direction.  A back-gate voltage $V_g$ is applied at the bottom surface to control the Fermi level.
The Hamiltonian of 2D MBT thin films reads~\cite{Sun2020} 
\begin{eqnarray} \label{Eq:MnBi2Te4}
H_{\textrm{TI}}(\boldsymbol{k}) &=& \epsilon_{0}(\boldsymbol{k}) +
\begin{bmatrix}
M(\mathbf{k}) & -iA_{1}\partial_z  & 0 & A_{2} k_{-} \\
-iA_{1}\partial_z & -M(\mathbf{k}) & A_{2} k_{-}  & 0 \\
0  & A_{2} k_{+} & M(\mathbf{k}) & iA_{1}\partial_z 
\\
A_{2} k_{+} & 0  & iA_{1}\partial_z & -M(\mathbf{k})
\end{bmatrix}
\nonumber\\
&&-e\phi(z)+H_X(z),
\end{eqnarray}
The translational invariance in the $x-y$ plane allows us to consider
a fixed in-plane wave vector $\boldsymbol{k}=(k_x,k_y)$ of magnitude $k=|\boldsymbol{k}|$. And $k_{\pm}=k_{x} \pm i k_{y}$, $\epsilon_{0}(\boldsymbol{k})=  C_{0}  -D_{1}\partial_{z}^{2} + D_{2}(k_{x}^{2}+k_{y}^{2})$ and $M(\boldsymbol{k})=M_{0} - B_{1}\partial_{z}^{2} +B_{2}(k_{x}^{2}+k_{y}^{2}) $.  In our calculations, $C_0$, $D_i$, $M_0$, $B_i$ and $A_i$ with $i=1,2$, are model parameters adopted from \emph{ab initio} calculations, see Appendix~\ref{Appendix A}.  $H_X$ is the spatial profile of the exchange field in the antiferromagnetic MBT.
For simplicity, we consider $H_X$ in terms of the sinusoidal function, which takes the form~\cite{Sun2020} 
\begin{equation}\label{eq-hx}
H_{X}(z) = -m_0\sin{\left(\frac{\pi}{d}z\right)}s_z\sigma_0,
\end{equation}
where $m_0$ is the amplitude of the intralayer ferromagnetic order, $d$ is the thickness of a septuple layer (SL), $s_j$ and $\sigma_j$ ($j=0,x,y,z$) are Pauli matrix acting in spin and orbital space, respectively.
$\phi(z)$ is electrostatic potential,  which is obtained by Schr\"odinger-Poisson (SP) method~\cite{Vuik-IOP-2016,Antipov2018,Mikkelsen2018}.

\begin{figure*}[t]
\centerline{\includegraphics[width=2\columnwidth]{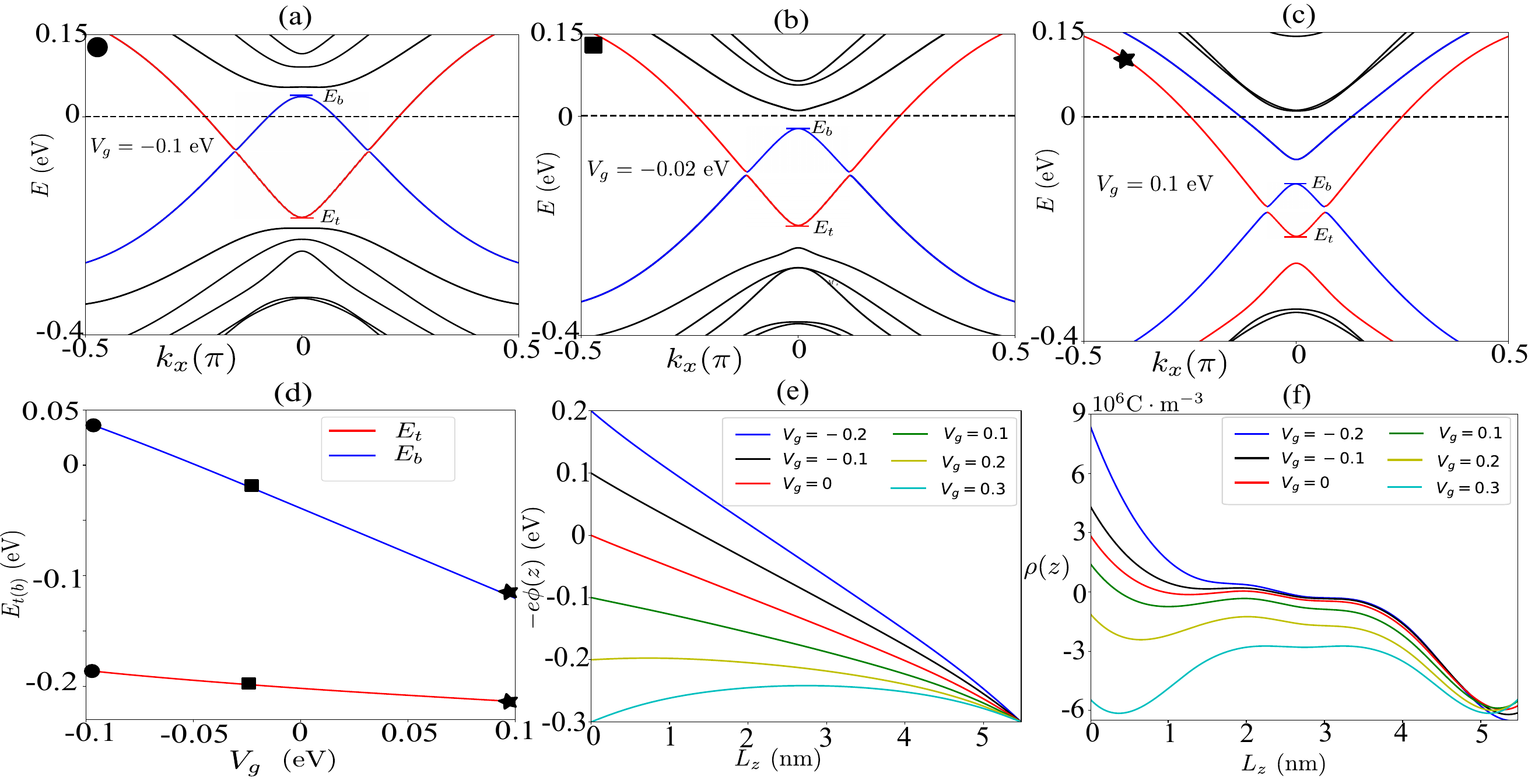}}
\caption{(a)-(c) Energy bands of MBT with different gate voltage $V_g$. 
The blue, red, and black curves correspond to BSSs, TSSs, and bulk states, respectively.
(d) Gate voltage dependence of the  eigen-energies of TSSs and BSSs, see $E_b$ and $E_t$ in panels (a)-(c). Note the three different corresponding markers. (e) The distribution of electrostatic potential energy $-e\phi(z)$.  
The right side is fixed by the band bending strength between MBT and SC. 
(f) The distribution of carrier charge density $\rho(z)$.}
\label{fig-band}
\end{figure*}

In order to obtain $\phi(z)$ self-consistently, we firstly set a initial potential $\phi_0(z)$ into the Hamiltonian $H_{\textrm{TI}}$. In our calculations, we choose $\phi_0(z)$ to be a constant function $\phi_0(z) = V_g$. Then we solve the Schr\"odinger-Poisson,  
\begin{equation}\label{supp-sp-s}
H_{\textrm{TI}}(\boldsymbol{k},\phi_0(z))\Psi_{n,\boldsymbol{k}}(z) =E_{n,\boldsymbol{k}}\Psi_{n,\boldsymbol{k}}(z),
\end{equation}
producing a set of eigenenergies $E_{n,\boldsymbol{k}}$, and a corresponding set of eigenstates $\Psi_{n,\boldsymbol{k}}(z)$. $n$ is the index of the transverse eigen-functions.  
Since the superconductor is typically metallic and screens electric fields perfectly~\cite{Mikkelsen2018}, 
we solve the Schr\"odinger equation only in the MBT region.
It means that we treat the SC only as
a boundary condition with a band offset $W$ at the interface between the MBT
and SC.
The charge density with the potential profile $\phi_0(z)$ is obtained by integrating over the occupied eigenstates and minus the density stems from the whole valence band $\rho_{\textrm{val}}(z)$
\begin{equation}\label{supp-sp-c}
\rho_1(z) = \frac{-e}{2\pi}\int_0^{\infty}\bigl[\sum_{n}|\Psi_{n,\mathbf{k}}(z)|^2f_{T}(E_{n,\mathbf{k}})-\rho_{\textrm{val}}(z)\bigr]kdk,
\end{equation}
where $f_{T}(E_{n,\mathbf{k}}) = 1/(e^{E_{n,\mathbf{k}}/T}+1)$ is Fermi distribution. Because $H_{\textrm{TI}}$ is a four band $k \cdot p$ Hamiltonian, we choose $\rho_{\textrm{val}}(z) = 2$. A new potential $\phi_1(z)$ is obtained by solving the Poisson equation
\begin{equation}\label{supp-sp-p}
\frac{d^2\phi_1(z)}{dz^2} = -\frac{\rho_1(z)}{\epsilon_r \epsilon_0},
\end{equation}
where $\epsilon_r$ denotes the dielectric constant of the MBT. As discussed previously, the boundary conditions of Eq.~\eqref{supp-sp-p} are $\phi(0)=V_g$ and $\phi(L_z)=W$.
Usually, $\phi_1(z)$ is not consistent with the initial potential $\phi_0(z)$. The error is defined as 
\begin{equation}\label{supp-sp-error}
\sigma_1 = \frac{\sum_m\left[\phi_1(z_m)-\phi_{0}(z_m)\right]^2}{N_m},
\end{equation}
where subscript of $\sigma_1$ represents the number of iterations. $m$ is the site index and $N_m$ is the number of sites.

The SP problem requires a self-consistent solution of two iterative equations Eq.~\eqref{supp-sp-s} and Eq.~\eqref{supp-sp-p}  until the error of $i$-th iteration $\sigma_i$ is smaller than the critical value $\sigma_c$. And the output $\phi_i(z)$ is the final self-consistent potential. The most straightforward iterative method is to replace the potential in Eq.~\eqref{supp-sp-s} directly with the newly obtained potential in Eq.~\eqref{supp-sp-p}. However, this usually leads to divergence of the iterations, and requires the suitable choice of initial potential $\phi_0(z)$. Thus, we employ a mixing scheme~\cite{Mikkelsen2018}, where the input potential used in each iteration is a mixing of
the input and output  potential of the previous
iteration:
\begin{equation}\label{supp-sp-interation}
\phi_i^{\textrm{in}}(z) = \kappa \phi_{i-1}^{\textrm{out}}(z) +(1-\kappa)\phi_{i-1}^{\textrm{in}}(z).
\end{equation}
e set $\kappa = 0.1$ and $\sigma_c = 10^{-8}$ eV in our calculations. 
In Fig.~\ref{supp-SP} we show an iterative procedure when we calculate the potential with $V_g = 0.2$ eV. As shown in Fig~\ref{supp-SP}(a), the iteration error decreases sharply as the number of iterations increase. The convergence of the potential [Fig~\ref{supp-SP}(b)] and charge density [Fig~\ref{supp-SP}(c)] occurs when the iterations number $n>50$ with the error $\sigma < 10^{-7}$ eV, see the black solid and dash lines.

\begin{figure*}[!htb]
\centerline{\includegraphics[width=2\columnwidth]{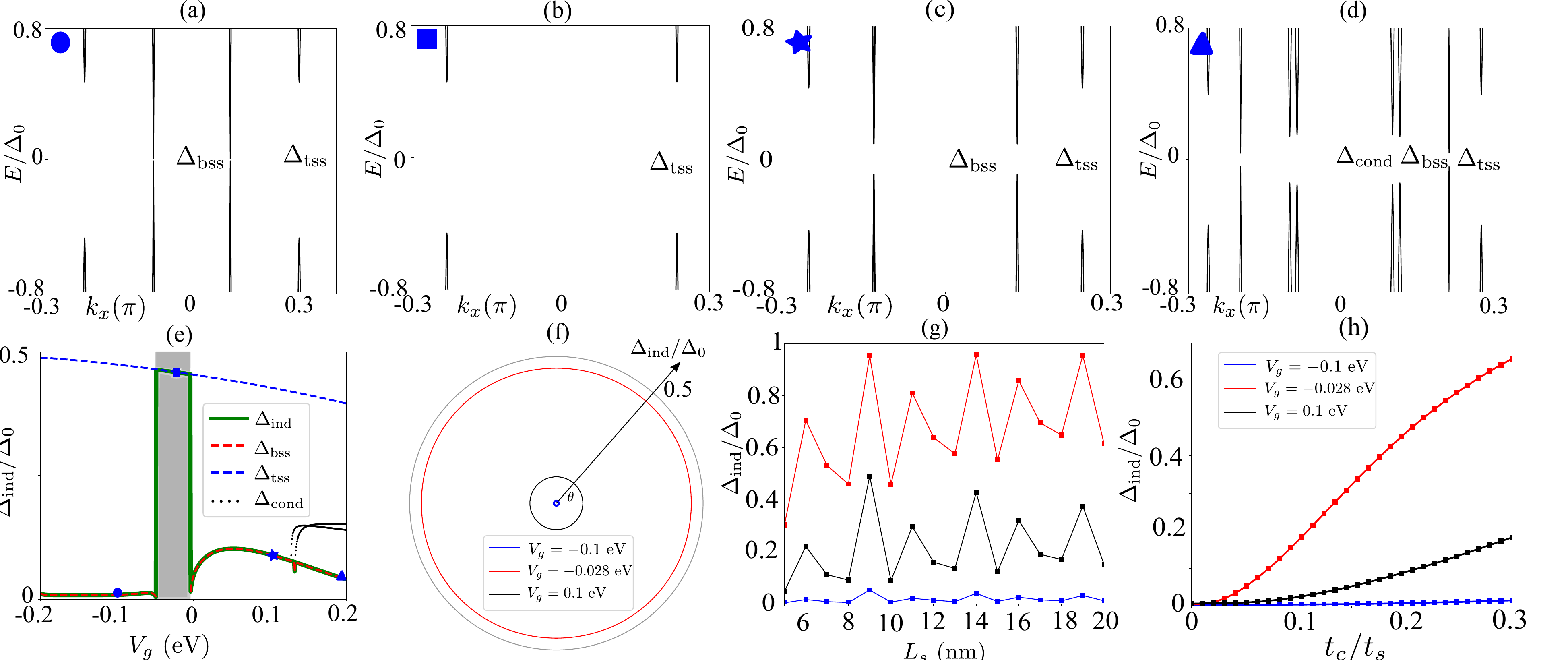}}
\caption{ (a)-(d) The energy bands of MTB-SC system with four typical gate voltages $V_g$ as marked in panel (e). (e) $\Delta_{\textrm{bss}}$, $\Delta_{\textrm{tss}}$ and $\Delta_{\textrm{cond}}$ represents the induced gap of BSSs, TSSs and conduction bands ($N=4$).   $\Delta_{\textrm{ind}}$ is the minimum of the gaps of all occupied states. The Chern number $\mathcal{C} = 1$ in the gray region, otherwise $\mathcal{C} = 0$. (f) The polar representation of the SC gap $\Delta_{\textrm{ind}}$ along the whole Fermi surface of MBT.  The radial length represents the amplitude of $\Delta_{\textrm{ind}}$. (g)  $\Delta_{\textrm{ind}}$ as a function of the thickness of superconductor $L_s$. The size of the induced gap oscillates with a period set
by the Fermi wavelength of the superconductor $\lambda_s$ (for our parameters $\lambda_s = 1.94$ nm), which is consistent with results in Ref.~\cite{reeg-prb-2018,Legg-RPB-2022}. (h)  $\Delta_{\textrm{ind}}$ as a function of coupling strength $t_c$. $t_s = \hbar^{2}/2 m_s a_s^2$ is the hopping magnitude in the superconductor. In panels (f)-(h), the blue, black and red curves correspond to the cases with gate voltage $V_g = -0.1$ eV, $V_g = -0.028$ eV and $V_g = 0.1$ eV respectively. }
\label{fig-hyrbidband}
\end{figure*}

The chemical potential of MBT can be obtained from $\phi(z)$ solution with different gate voltage $V_g$, as shown in Figs.~\ref{fig-band}(a)-(c).  
Here we choose the septuple layer number of MBT $N=4$ (with the full thickness $L_0 = Nd$). The inhomogeneous electrostatic potential breaks inversion symmetry, which lifts the degeneracy of the surface states.
Obviously, the bottom surface states (BSSs) and the top surface states (TSSs) have totally different electrostatic environments because they couple to back-gate and SC, respectively. Therefore, the two surface states have different tunability with the change of gate voltage $V_g$. The Fermi level of TSSs can be well controlled by the different gate voltage, see Figs.~\ref{fig-band}(a)-(c).
However, the magnetic Dirac gap of TSSs is always below the Fermi level during the gate control. 
In Fig.~\ref{fig-band}(d), we calculate the eigen-energies of TSSs and BSSs at $k_x = 0$ (see $E_t$ and $E_b$ labeled in Figs.~\ref{fig-band} (a)-(c)) as a function of gate voltage $V_g$.  
Note that the Fermi level of TSSs (red line) is nearly unaffected by the change of $V_g$.
The different tunability between BSSs and TSSs stems from the non-uniform distribution of the electrostatic potential in MBT. As shown in  Fig.~\ref{fig-band}(e), the potential energy $-e\phi(z)$ at the SC-MBT boundary (right side) is fixed at $W = -0.3$ eV~\cite{rmann2022}, which is the band bending strength between MBT and SC (see Appendix~\ref{Appendix B} for details).
Nevertheless, the potential energy close to the MBT-substrate boundary (left side) varies with the gate voltage. Because TSSs distribute locally near the interface between MBT and SC, the tunability is greatly limited  by the band bending effect.
This constraint about the chemical potential tunability also manifest in the charge density distribution [Fig.~\ref{fig-band}(f)]. The type of carrier near the left side is electron (hole) when $V_g$ is positive (negative). While the carrier near the right side is nearly unchanged with $V_g$.
Actually, these results, which cannot be obtained in previous minimal models~\cite{WangJi2016,Peng2019PRB,Yan-PRB-2019,He-CP-2019,YZB-PRB-2019,Sun2020,Zhang-PRB-2021,Hogl-PRB-2020,Luo-PRB-2021,Zhang-PRL-2021}, could highly narrow the regions of parameter for achieving CMMs.

\section{Superconducting proximity effect}\label{proximity}
When the superconducting shell is considered,
the Bogoliubov-de Gennes (BdG) Hamiltonian of MBT-SC hybrid system takes the form
\begin{eqnarray}\label{Hbdg}
H_{\textrm{BdG}} = \begin{pmatrix}
	H_{\rm TI}+H_{s}+H_t & is_y\Delta(z)  \\
		-is_y\Delta(z)  & -(H_{\rm TI}+H_{s}+H_t)^{*}
	\end{pmatrix}
\end{eqnarray}
We include an $s$-wave pairing potential only in the SC part, i.e., $\Delta(z) = \Delta_0$ for $z>L_{0}$, and $\Delta(z)=0$ for $z<L_0$. The normal state of the SC has the form $H_{s} = \frac{\hbar^2 \boldsymbol{k}^2}{2m_e}-\mu_s$, where $\mu_s$ is the chemical potential and the effective mass $m_s$ is taken to be infinite in the direction parallel to the interface~\cite{Vaitiekenas-science-2020}. It is noted that the calculated self-consistent electrostatic potential $\phi(z)$ is included in $H_{\rm TI}$.
The coupling between the MBT and SC at the interface takes the from~\cite{Legg2022}
\begin{equation}\label{hop}
    H_{t} = \sum_{\langle z,z^{'} \rangle}\bigl[-t_c c_{z,k}^{\dagger}d_{z^{'},k}+\textrm{H.c.}
    \bigr],
\end{equation}
where $\langle z,z^{'} \rangle$ denotes the hopping between the nearest sites. $t_c$ is the coupling strength. The operator $c_{z,k} (d_{z,k})$ annihilates a state of momentum $k$ at site $z$ within the MBT (SC).

\begin{figure*}[!htb]
\centerline{\includegraphics[width=2\columnwidth]{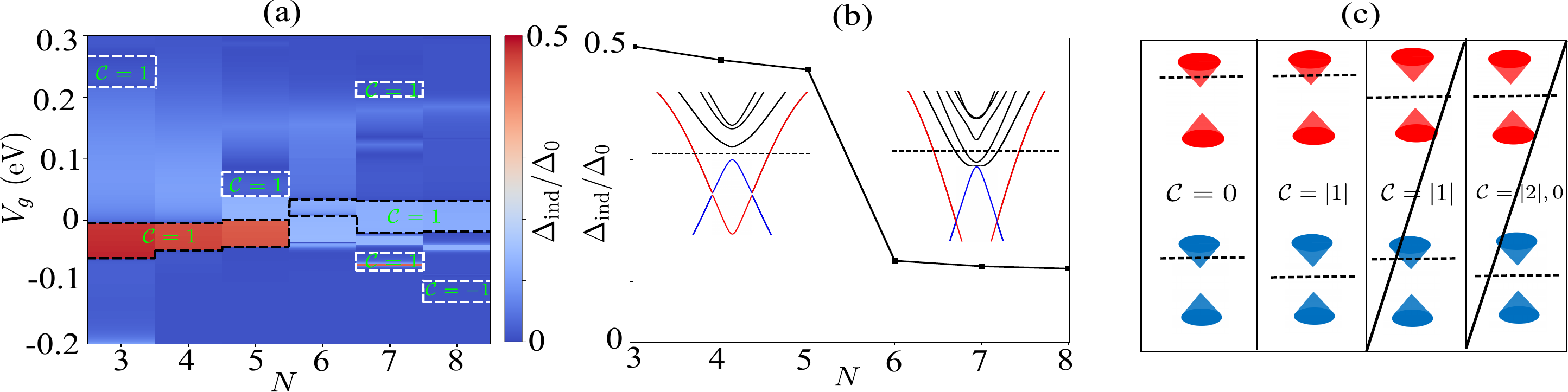}}
\caption{(a) Topological phase diagram as a function of gate voltage $V_g$ and the layers of MBT $N$.  Topological regions which stem from surface states and bulk states are enclosed by the black and white dash lines, respectively.  (b) Thickness dependence of the induced gap. For each layer number $N$, we calculate the largest SC gap in the topological region enclosed by black dash lines in panel (a). When $N$ increases up to six SLs, the SC gap decrease drastically because the Fermi level crosses the conduction bands. Inset: The electron band  of MBT when $N=4$ (left) and $N=8$ (right).  (c) The Fermi level of TSSs  can not be tuned in the magnetic Dirac gap. This narrows the topological regions in contrast to those predicted in previous works, as indicated by the slash.  }
\label{fig-phase-diagram}
\end{figure*}

The induced SC gap of MBT is highly dependent on the types of the bands crossing the Fermi level [Fig.~\ref{fig-hyrbidband}(e)]. We use $\Delta_{\textrm{bss}}$, $\Delta_{\textrm{tss}}$ and $\Delta_{\textrm{cond}}$  to represents the induced gap of BSSs, TSSs and conduction bands. The induced gap of MBT is defined as the minimum of all the gaps $\Delta_{\textrm{ind}} = \textrm{min}\{\Delta_{\textrm{bss}}, \Delta_{\textrm{tss}},\Delta_{\textrm{cond}}\}$.
In Figs.~\ref{fig-hyrbidband}(a)-(d), we calculate the SC bands with four typical $V_g$, as marked in Fig.~\ref{fig-hyrbidband}(e).   When  MBT is bulk insulating and Fermi level only crosses both BSSs and TSSs [Fig.~\ref{fig-hyrbidband}(a)], TSSs opens a finite SC gap with $\Delta_{\textrm{tss}}/\Delta_0 \approx 0.462$. 
Arguably, superconductivity at the BSSs may be strongly suppressed (see a recent experiments~\cite{Hlevyack2020}) because of the very short penetration depth, about 1.62 nm~\cite{Hlevyack2020,Sun2020,Shikin-Sci_report-2020}. In our calculations, the thickness of MBT is 5.48 nm (4 SLs). 
Thus, the suppression of superconductivity on BSSs limits the SC proximity gap, except for the region where the Fermi level is tuned in the magnetic gap of BSSs [Fig.~\ref{fig-hyrbidband}(b)]. In this case, MBT has largest induced gap because Fermi level only cross TSSs [Fig.~\ref{fig-band}(b)], and $\Delta_{\textrm{ind}}$ is  dominated by $\Delta_{\textrm{tss}}$. 
When the Fermi level moves toward the bottom of conduction bands,  $\Delta_{\textrm{bss}}$ gradually increases because of the increase of penetration depth [Fig.~\ref{fig-hyrbidband}(c)]. 
When Fermi level crosses the conduction bands [Fig.~\ref{fig-hyrbidband}(d)], $\Delta_{\textrm{bss}}$ is still suppressed. 
This is because the positive gate-induced electrostatic potential will change the electrons’ confinement, and pull electron density away from the interface between MBT and SC~\cite{Antipov2018}. This in turn strongly suppresses the SC proximity effect for states in MBT [Fig.~\ref{fig-hyrbidband}(e)].

In our calculations, $\Delta_{\textrm{ind}}$ is obtained from the superconducting band with $k_y = 0$ because it is isotropic. In Fig.~\ref{fig-hyrbidband}(f), we plot the polar representation of $\Delta_{\textrm{ind}}$ along the whole Fermi surface. The blue, black, and red curves correspond to the cases with different gate voltage $V_g$. And the radial length represents the amplitude of $\Delta_{\textrm{ind}}$ along the Fermi surface. Clearly, $\Delta_{\textrm{ind}}$ is isotropic because of the circle shape. On the hand, the SC gap also depends on the thickness of the superconductor $L_s$. In Fig.~\ref{fig-hyrbidband}(g), we calculate  $\Delta_{\textrm{ind}}$ as a function of $L_s$. We find that the size of the induced gap oscillates with a period set
by the Fermi wavelength of the superconductor $\lambda_s$ (for our parameters $\lambda_s = 1.94$ nm), which is consistent with results in Ref.~\cite{reeg-prb-2018,Legg-RPB-2022}. Another significant parameter affecting the proximity effect is the coupling strength $t_c$ between the MBT and SC [Fig.~\ref{fig-hyrbidband}(h)]. Certainly, the SC gap gradually increase with the increases of $t_c$. 
Importantly, our results don't change qualitatively when varying the amplitude of $t_c$ and $L_s$. 
The  induced SC gap of MBT is highly dependent on the gate voltage $V_g$, i.e., the type of the bands crossing the Fermi level. MBT has the largest SC gap when it is in the chiral topological superconductor phase (red curves in Figs.~\ref{fig-hyrbidband}(f)-(h)). When MBT is in the trivial phase, 
the suppression of
superconductivity on BSSs limits the SC proximity gap (blue and black curves in Figs.~\ref{fig-hyrbidband}(f)-(h)). 
These results about the gate tunability cannot be captured by previous minimal models.

\section{chiral Majorana mode}\label{chiral-MBS}
As discussed above, we mainly have two constraints that limit the realization of CMMs. Firstly, the tunability of TSSs is greatly limited by the band bending
effect. Secondly, BSSs exhibit a giant attenuation of surface superconductivity. 
Nevertheless, we demonstrate that CMMs can still be achieved in a reasonable range of experimental parameters. 
The key requirement for realizing CMMs is to achieve superconductivity and magnetization on TSSs and BSSs, respectively. 
As shown in Figs.~\ref{fig-band}(a)-(c), the magnetic Dirac gap of TSSs is always much below the Fermi level during the gate tuning. This fact protects the induced superconductivity of TSSs from the destruction of the magnetization~\cite{chen-prr-2021}.
Thus, the key point for achieving CMMs is to tune the Fermi level into the magnetic gap of BSSs, which is enabled by the high tunability [Figs.~\ref{fig-band}(d)]. 
We further calculate the Chern number $\mathcal{C}$ (see Appendix~\ref{Appendix C}) as a function of gate voltage. We have $\mathcal{C} = 1$ in the gray region of Fig.~\ref{fig-phase-diagram}(e) where the Fermi level locates in the magnetic gap of BSSs, otherwise  $\mathcal{C} = 0$.

Fig.~\ref{fig-phase-diagram}(a) shows the topological phase diagram as a function of the two experimentally
relevant and tunable quantities -- gate voltage $V_g$ and the layer number of MBT $N$, rather than more phenomenological
parameters. 
Note that most of the topological regions, which are enclosed by the black dash lines, are concentrated in the range of $V_g \in (-0.06~0.03)$ eV. 
When the thickness of MBT increases up to six SLs,  the superconducting gap in the topological regions decreases drastically because the Fermi level also crosses the conduction bands [Fig.~\ref{fig-phase-diagram}(b)].  
Another remarkable result is that 
we also have additional topological regions stemming from the bulk states of MBT (Appendix~\ref{Appendix D}), which are enclosed by the white dash lines in Fig.~\ref{fig-phase-diagram}(a).
The formation of these topological regions originates from two major effects on bulk states: induced finite spin-orbital coupling due to the applied electric field and the magnetization effects.
Because of the antiferromagnetic structure of MBT,
these usually occur when the layer number is odd or the gate voltage is negative. Nevertheless, the induced superconducting gaps in these topological regions are very small, which is not favorable for achieving robust CMMs.
Notably, the obtained topological regions in Fig~\ref{fig-phase-diagram}(a) are greatly narrowed compared with those predicted in previous works (Appendix~\ref{Appendix E}). This is because the Fermi level of TSSs can not be tuned into the magnetic Dirac gap, i.e., CMMs with Chern number $\mathcal{C}=2$ can not be realized, as illustrated in Fig~\ref{fig-phase-diagram}(c).
It is noted that the phase diagram [Fig.~\ref{fig-phase-diagram}(a)]  does not change qualitatively as long as $W$ is not very small. Otherwise, the Fermi level of TSSs can also be tuned by the gate voltage, and additional topological regions stemming from TSSs will arise (Appendix~\ref{Appendix B} ).

\begin{figure}[!htb]
\centerline{\includegraphics[width=1\columnwidth]{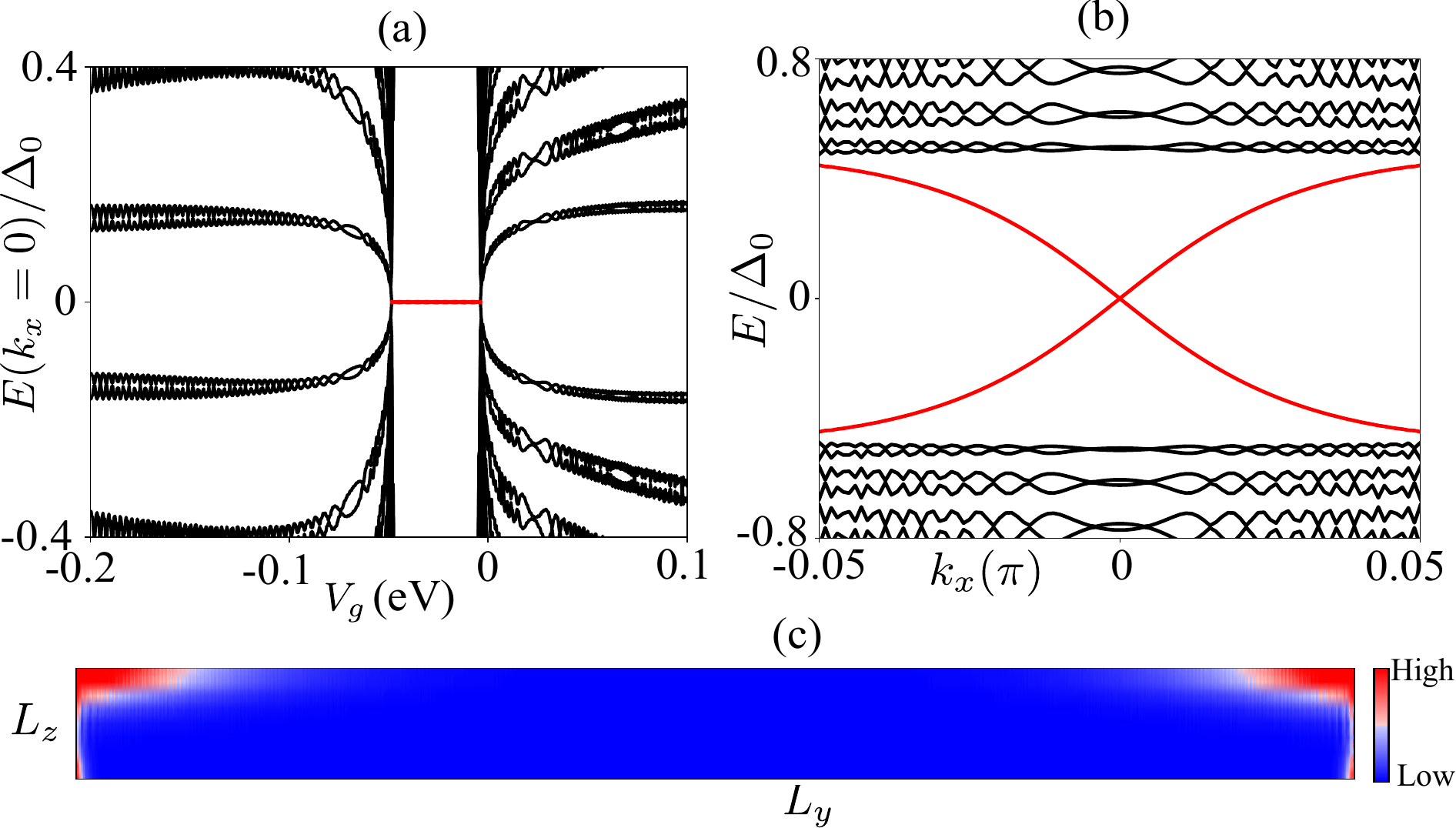}}
\caption{ (a) The eigen-energies at $k_x = 0$ as a function of gate voltage $V_g$. The red curves correspond to the CMMs.
(b) The spectrum shows that CMMs (red curves) appear in the SC gap.  (c) The
distribution of local density of states of CMMs at $k_x = 0$ in the
$y-z$ cross section of MBT.
The SC part of MBT-SC is not shown. We choose $N=4$, $V_g = -0.028 $ eV and $L_y = 1.5~\upmu$m.}
\label{Fig-chiral-MBS}
\end{figure}

To further confirm the system is exactly in the topological phase under such conditions,  we consider the MBT-SC system
with open boundary condition in the $y$ direction.  In Fig.~\ref{Fig-chiral-MBS}(a), we calculate the eigenenergy with $k_x = 0$ as a function of $V_g$. 
The gap closes when $V_g$ approaches -0.048 eV. Then a pair of zero modes emerges in the gap, which is the crossing point of the two CMBSs at $k_x = 0$ [Fig.~\ref{Fig-chiral-MBS}(b)].  The distribution of local density of CMMs at $k_x = 0$ in the $y$-$z$ cross section (the top SC part is not shown) is given in Fig.~\ref{Fig-chiral-MBS}(c). As expected, CMMs mainly distribute in the two edges of the MBT-SC slab and gradually decay into the bulk.

\section{Conclusion and discussion}\label{Conclusion}
We consider a two dimensional MBT thin film in proximity to an s-wave SC.  Beyond the well-known minimal models, we calculate the electrostatic potential self-consistently in a Schr\"odinger-Poisson scheme. We find that the band bending effect at MBT-SC interface 
severely restricts the tunability of top surface states, and the corresponding magnetic Dirac gap is always below the Fermi level during the gate tuning. Moreover, we find that the induced SC gap of MBT is highly dependent on the types of the bands crossing the Fermi level. Arguably, superconductivity at the BSSs may be strongly suppressed especially when bulk is insulating. 
 These results, which cannot be obtained in previous minimal models, could highly narrow the regions of parameter for achieving CMMs. 
 Nevertheless, we demonstrate that the CMMs can still be realized via the control of the gate voltage. The key point is to tune the Fermi level of the bottom surface state into the magnetic Dirac gap.
Our method provides a more accurate prediction about the topological phase and device control capability. This is in stark contrast to those previous theoretical work. 

In this work, we regard MBT as an infinite 2D system and only consider the inhomogeneity of the potential in the $z$ direction. 
This approximation is reasonable because the size of the MBT is usually very large, about hundreds of nanometers. And most of the wavefunctions of the TI surface states are localized on the top and bottom surfaces.
Thus, the electrostatic properties of surface states are barely affected by the potential spikes at the edges.
Apart from the Schr\"odinger-Poisson method,  another self-consistent method which is called the Thomas-Fermi method, has also been widely used in Rashba semiconductors~\cite{Mikkelsen2018}. However, the Thomas-Fermi approximation relies on the
assumption that the electronic charge density is given by the standard result for a homogeneous 3D electron gas. Thus, it is not appropriate in the TI system because of the existence of topological surface states.  Compared with bulk states, surface states are more concentrated near the interface, so they are more sensitive to band bending. Thus, our result that topological surface states near the superconductor do not respond to gating does not apply to the bulk states in the nanowire system. Although the disorder defects are not considered in this work, it was still a very important topic. Various novel phenomena were proposed to exist in disordered QAH systems~\cite{Groth2009-PRL,Huang2018a-PRB,Lian2018-PRB}. More calculations about different types of disorder effects in the QAH-SC system will be considered in the future.

Apart from the QAHI system, the electrostatic effects also exist in other TI-SC hybrid systems, such 
as MZMs in the vortex of SC-TI~\cite{Fu2008} and TI nanowire~\cite{Franz-PRB-2011,Legg-prb-2011,Muenning2021}. Growing TI film on SC substrate will induce charge doping from the SC to the TI, which shifts the Fermi level into the TI conduction band~\cite{Xu2014,Xu-PRL-2015,rmann2022}. Thus, MZMs only exist when TI film is enough thick, at least 3 quintuple layers as they found in Ref.~\cite{Xu-PRL-2015}. As for TI nanowire, it was proposed that in-homogeneous potential breaks the inversion symmetry, which enhances the sub-band splitting of TI states required for the realization of topological superconductivity~\cite{Legg-prb-2011}.

\begin{acknowledgements}
{\em Acknowledgments - } Authors thank Yayu Wang, Yang Feng, and Gu Zhang for helpful discussions. This work was supported by the Innovation Program for Quantum Science and Technology (Grant No.~2021ZD0302400, No.~2021ZD0302700), the National Natural Science Foundation of China (Grants No.~11974198, No.~12004040,No.~12074133), and Tsinghua University Initiative Scientific Research Program.
\end{acknowledgements}

\appendix
\section{Parameters used in this work}\label{Appendix A}
\setcounter{equation}{0}
\renewcommand\theequation{A\arabic{equation}}
The parameters of the $k\cdot p$ Hamiltonian of MBT in Eq.~\eqref{Eq:MnBi2Te4} are adopted from \emph{ab initio} calculations~\cite{Zhang2019PRL}: $C_0 = -0.05~\textrm{eV}$, $M_0 = -0.117~\textrm{eV}$, $D_1 = 2.72~\textrm{eV \AA$^2$}$, $D_2 = 1.2~\textrm{eV \AA$^2$}$, $B_1 = 11.9~\textrm{eV \AA$^2$}$, $B_2 = 9.40~\textrm{eV \AA$^2$}$, $A_1 = 2.7~\textrm{eV \AA}$, $A_2 = 3.2~\textrm{eV \AA}$. The other parameters used in this work are given in Table~\ref{Table:parameters}. $a_{x,y,z}$ is the lattice constant of MBT in the tight-binding calculations. In our calculations, the choice of the superconducting (SC) materiel is NbSe$_2$, which has been widely used in experiments~\cite{Xu2014,Wang-science-2012,Xu-PRL-2015,Sun-PRL-2016}.  $a_s$ is the lattice constant of SC. Because the dielectric constant of MBT has not been studied experimentally, we set it equal to the value of Bi$_2$Se$_3$. The details about band bending are in  Appendix~\ref{Appendix B}. 

\begin{table}[!ht]
\caption{Parameters used for the calculations in this work. \label{Table:parameters}}
\begin{tabularx}{\columnwidth}{X >{\centering}X >{\centering}X>{\centering}X >{\centering\arraybackslash}X}
\hline
\hline
 $m_0$ & $d$ & $\Delta_0$ & $\mu_s$ & $L_{s}$\\
0.1 eV~\cite{Shikin-Sci_report-2020} & 1.37 nm~\cite{Gong_2019} & 1.5 meV~\cite{Clayman-1972} & 0.4 eV~\cite{Yokoya-science-2001} & 10 nm  \\ \hline
 $\epsilon_r$ & $a_s$ & $a_{x(y)}$ & $a_z$ & $m_s$ \\
 25~\cite{Stordeur-1992}  & 0.4 nm  & 1 nm & 0.7 nm & $m_e$ \\\hline
 $W$ & $t_c$ & & & \\
 0.3 eV~\cite{Xu2014} & 0.05 eV & & &\\
\hline
\hline
\end{tabularx}
\end{table}

\section{Effect of band bending strength}\label{Appendix B}
\setcounter{equation}{0}
\renewcommand\theequation{B\arabic{equation}}

\begin{figure*}[!htb]
\centerline{\includegraphics[width=2\columnwidth]{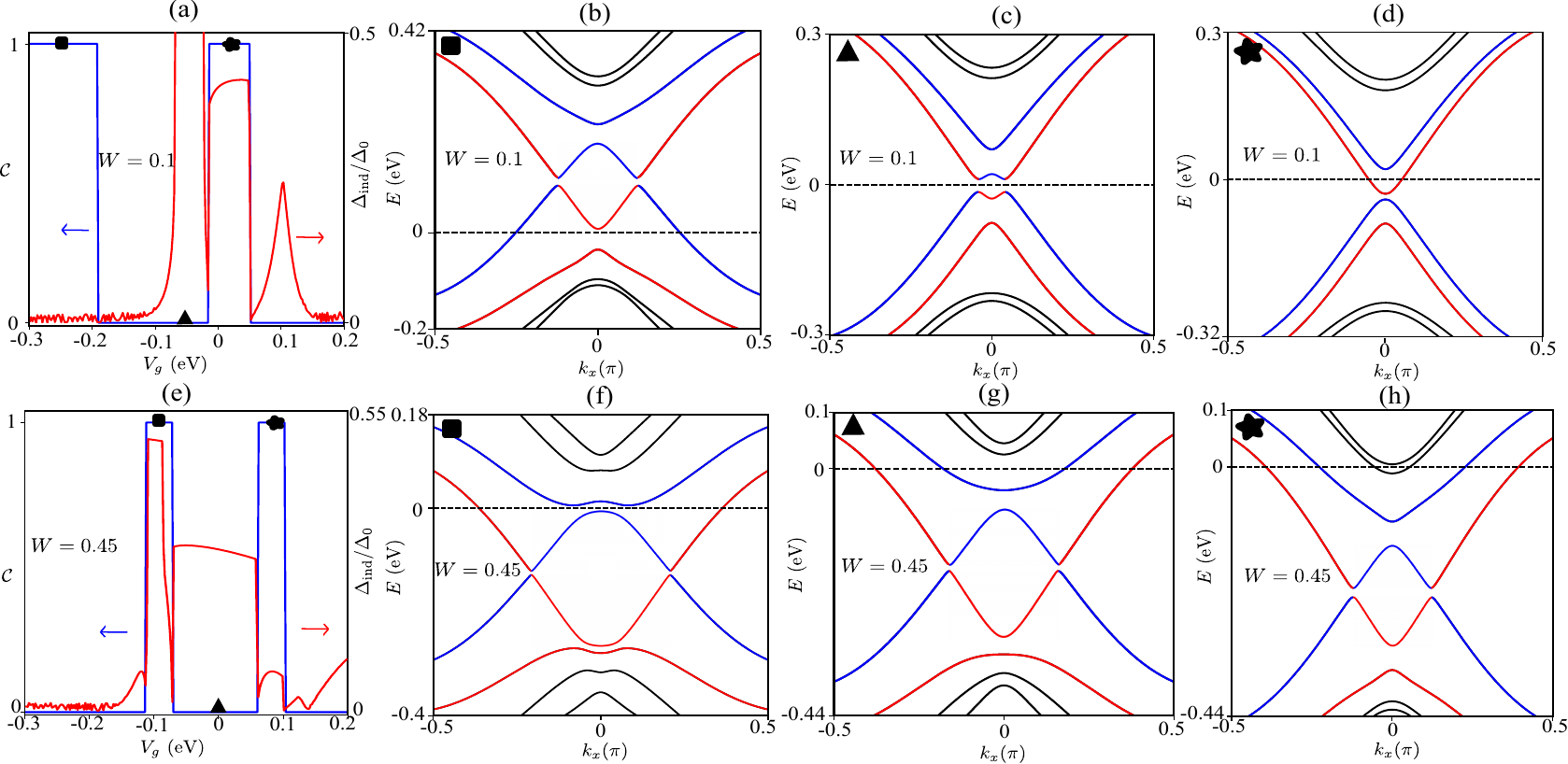}}
\caption{(a) The Chern number $\mathcal{C}$ (blue curves) and induced SC gap $\Delta_{\textrm{ind}}$ (red curves) as a function of gate voltage $V_g$ when the band bending strength $W = 0.1$ eV.  (b)-(d) The energy band structure of MBT. The blue, red, and black curves correspond to BSSs, TSSs, and bulk states, respectively. We choose the gate voltage of panels (b)-(d) as marked in panel (a).  
In panels (b) and (d), the Fermi level is tuned in the magnetic gap of the BSSs and TSSs, respectively. In panel (c), the Fermi level is tuned in the trivial gap stemming from the coupling between BSSs and TSSs.
(d)-(f) The cases when $W = 0.45$ eV. In panels (f) and (h), the Fermi level is tuned in the magnetic gap of the BSSs and first lowest bulk band, respectively. 
Due to the large band bending effect, the Fermi level can not be tuned in the magnetic gap of TSSs. The layers number of MBT $N$ is fixed to 3 in all the panels. }
\label{supp-band-bending}
\end{figure*}

\begin{figure*}[ht]
\centerline{\includegraphics[width=2\columnwidth]{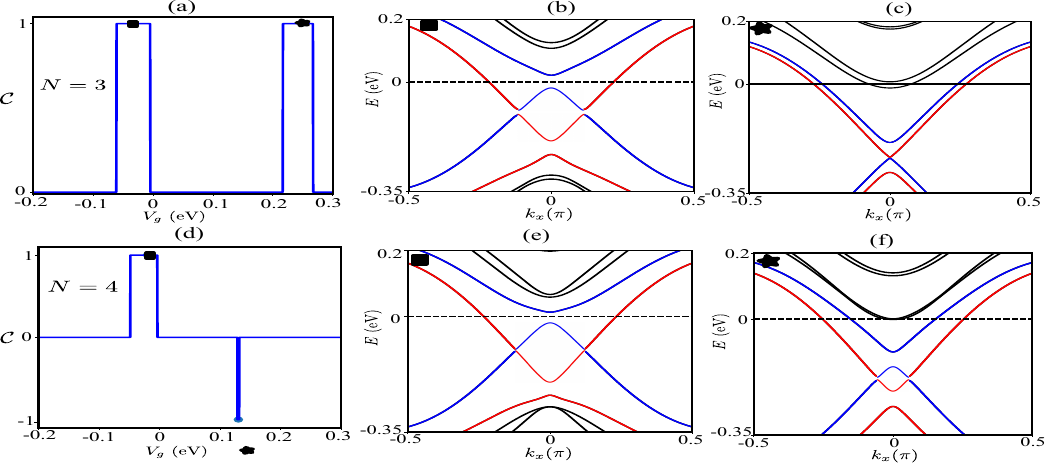}}
\caption{(a) The Chern number $\mathcal{C}$ as a function of gate voltage $V_g$ when the layer number of MBT $N=3$. There exist two topological regions with $\mathcal{C} = 1$, which stem from surface states and bulk states respectively. (b)-(c) The band structure of MBT. The blue, red, and black curves correspond to BSSs, TSSs, and bulk states, respectively. We choose the gate voltage of panels (b) and (c) as marked in panel (a). 
In panels (b) and (c), the Fermi level is tuned in the magnetic gap of the BSSs and first lowest bulk band, respectively. (d)-(f) The cases when $N = 4$. Due to the antiferromagnetic properties of MBT, the magnetization of bulk states is small when $N$ is even. So the corresponding topological region is small. In addition, the signs of the $C$ stem from surface states and bulk states are different. This is because the gate-induced electrostatic potential will confine the bulk states closer to the top surface of MBT. Thus, the BSSs and bulk states have opposite magnetization.   }
\label{supp-bulk-TSC}
\end{figure*}

The bend bending strength $W$ exists at the interface between MBT and SC because of their work function imbalance~\cite{KIEJNA-1996}.
In this section, we discuss the effect of band bending strength on our result. We consider two limiting cases: a small band bending strength with $W = 0.1$ eV and a very large value, $W = 0.45$ eV.

In Fig.~\ref{supp-band-bending}(a), we calculate the Chern number $\mathcal{C}$ (blue curves) and induced SC gap $\Delta_{\textrm{ind}}$ (blue curves) as a function of gate voltage $V_g$ when the band bending strength $W = 0.1$ eV. In Figs.~\ref{supp-band-bending}(b)-(d), we plot the energy bands of MBT with three typical different gate voltages as marked in Fig.~\ref{supp-band-bending}(a). Because of the small band bending strength, the Fermi level can be tuned in the magnetic gap of TSSs [Fig.~\ref{supp-band-bending}(b)]. Additional topological regions arise in this case, but the corresponding SC gap is very small because the superconductivity of BSSs is suppressed [Fig.~\ref{supp-band-bending}(a)]. When $W$ is very large [Figs.~\ref{supp-band-bending}(e)-(h)], the Fermi level can not be tuned in the magnetic gap of TSSs, which is consistent with the result in the main text. 

The exact value of  $W$  is unknown, depending on the choice of superconducting materials in the experiment. In addition, $W$ also depends on the plane of SC crystal~\cite{Lang-PRB-1971,Eastment_1973}, as well as the thickness of SC~\cite{Qi-APL-2007}. The work functions of MBT, Bi$_2$Te$_3$ and several SCs are given in Table~\ref{Table:work-functions}. The band offset between Bi$_2$Te$_3$ and NbSe$_2$ is about 0.15-0.2 eV according to the experiments in Ref.~\cite{Xu2014}. Obviously, the band banding strength in the MBT-NbSe$_2$ system is larger because of the smaller work function of MBT. In the main text, we set $W = 0.3$ eV. Nevertheless, our results do not change qualitatively as long as $W$ is not 
very small, see Fig~\ref{supp-band-bending}.

\begin{table}[!ht]
\caption{The work functions of MBT, Bi$_2$Te$_3$ and several SCs. \label{Table:work-functions}}
\begin{tabularx}{\columnwidth}{X >{\centering}X >{\centering}X>{\centering}X>{\centering}X >{\centering\arraybackslash}X}
\hline
\hline
 MnBi$_2$Te$_4$ &  Bi$_2$Te$_3$ & NbSe$_2$ & NbN & Al & Pb\\
4.0 \textrm{eV}~\cite{Akhgar2022} & 5.3 \textrm{eV}~\cite{Shih-PRL-2010} & 5.9 \textrm{eV}~\cite{Shimada-1994} & 4.7 \textrm{eV}~\cite{Gotoh2003} & 4.26 \textrm{eV}~\cite{Eastment_1973} & 4.25 \textrm{eV}~\cite{Lang-PRB-1971}\\
\hline
\hline
\end{tabularx}
\end{table}

\section{Numerical calculation of Chern number}\label{Appendix C}
\setcounter{equation}{0}
\renewcommand\theequation{C\arabic{equation}}

For the calculation of Chern number $\mathcal{C}$, we don't utilize the Hamiltonian $H_{\textrm{BdG}}$ (Eq.~\eqref{Hbdg} in the main text). This is because 
the bands which stem from the superconductor are trivial. And the dimension of the superconductor Hamiltonian is very large, which increased computational effort. Thus, we treat the superconductor as the self-energy $\Delta_{\textrm{ind}}$~\cite{Stanescu-PRB-2010} and consider the Hamiltonian
\begin{eqnarray}\label{supp-BdG}
H_{\textrm{BdG}} = \begin{pmatrix}
	H_{\rm TI} & is_y\Delta_{\textrm{ind}}  \\
		-is_y\Delta_{\textrm{ind}}  & -H_{\rm TI}^{*}
	\end{pmatrix}.
\end{eqnarray}
As discussed in the main text, $\Delta_{\textrm{ind}}$ is highly dependent on the gate voltage. And top surface states (TSSs), bottom surface states (BSSs), and bulk states have totally different induced SC gaps. In our calculations of Chern number, we set   $\Delta_{\textrm{ind}} = \Delta_{tss}$, where $\Delta_{tss}$ is the SC gap of TSSs. This is because the topological phase transition is mainly related to $\Delta_{tss}$~\cite{Qi2011}.
The Chern number $\mathcal{C}$ of Hamiltonian Eq.~\eqref{supp-BdG} is~\cite{Thouless-PRL-1982}
\begin{eqnarray}\label{supp-chern}
\mathcal{C}  = \frac{1}{2\pi}\int d^2 k F_{12}(k),
\end{eqnarray}
where the Berry connection $A_{\mu}(k)~(\mu = 1,2)$ and the associated field strength $F_{12}(k)$ are given by
\begin{eqnarray}\label{supp-berry}
A_{\mu}(k) &=& 
-i\sum_{E_n<0}\bra{\psi_n(k)}\partial_{\mu}\ket{\psi_n(k)}, \\ \nonumber
F_{12}(k) &=& \partial_1 A_2(k)-\partial_2 A_1(k),
\end{eqnarray}
where $\psi_n(k)$ and $E_n$ is the $n$th eigen-function and eigen-values of Eq.~\eqref{supp-BdG}. 
We calculate the Chern number $\mathcal{C}$ numerically according to the method proposed in Ref.~\cite{Fukui-JPSJ-2005}. 

We consider lattice points $k_l~(l=1,...,N_1N_2)$ on the two dimensional discrete Brillouin zone as
\begin{equation}
k_l = (k_{j_1},k_{j_2}),~~k_{j_{\mu}} = \frac{2\pi j_{\mu}}{N_{\mu}}, ~~(j_{\mu} = 0,...,N_{\mu}-1),
\end{equation}
The occupied multiplet of Hamiltonian $H_{\textrm{BdG}}$ is $\Psi = (\ket{\psi_1},...,\ket{\psi_M})$. The number of the occupied states $M$ is half of the dimension of $H_{\textrm{BdG}}$ because of the particle-hole symmetry. The $U(1)$ link variable is defined as  
\begin{equation}
U_{\mu}(k_l) = \frac{1}{\mathcal{N}_{\mu}(k_l)}\textrm{det}[\Psi^{\dagger}(k_l)\Psi(k_l+\hat{\mu})],
\end{equation}
where $\hat{\mu}$ is a vector in the direction $\mu$ with the magnitude $\frac{2\pi}{N_{\mu}}$, $\mathcal{N}_{\mu}(k_l) = |\textrm{det}[\Psi^{\dagger}(k_l)\Psi(k_l+\hat{\mu})]|$ is the normalization constant. Then the lattice field strength is 
\begin{equation}
F_{12}(k_l) = \textrm{ln}\left[U_1(k_l)U_2(k_l+\hat{1})U_1(k_l+\hat{2})^{-1}U_2(k_l)^{-1}\right].
\end{equation}
And the Chern number $\mathcal{C}$ is the summation of the lattice field $F_{12}$ 
\begin{equation}
\mathcal{C} = \frac{1}{2\pi i}\sum_{l}F_{12}(k_l).
\end{equation}

\section{Topological regions stem from bulk states}\label{Appendix D}
\setcounter{equation}{0}
\renewcommand\theequation{D\arabic{equation}}

 The applied electric field will induce finite spin-orbital coupling on bulk states. And topological regions will exist as long as the magnetization of bulk states is enough large, i.e., satisfying the topological phase transition condition $M_{n,\textrm{bulk}}^2 > \Delta_{n,\textrm{bulk}}^2+ \mu_{n,\textrm{bulk}}^2$. Here $M_{n,\textrm{bulk}} $ $\mu_{n,\textrm{bulk}}$, and $\Delta_{n,\textrm{bulk}}$ is the magnetization, chemical potential and induced SC gap of the $n$th bulk bands, respectively. 

In Fig.~\ref{supp-bulk-TSC}(a), we calculate the  Chern number $\mathcal{C}$ as a function of gate voltage $V_g$ when the layer number of MBT $N=3$. Note that there exist two topological regions with $\mathcal{C} = 1$, which stem from surface states and bulk states respectively. To see it more clearly,  we choose the two gate voltage as marked in Fig.~\ref{supp-bulk-TSC}(a), and calculate the corresponding band structure of MBT [Fig.~\ref{supp-bulk-TSC}(b)(c)]. The blue, red, and black curves correspond to BSSs, TSSs, and bulk states respectively.   In Fig.~\ref{supp-bulk-TSC}(b) and (c), the Fermi level is tuned in the magnetic gap of BSSs and the first lowest bulk band, respectively. This indicates that the nonzero Chern number in these two cases stem from surface states and bulk states, respectively.
Due to the antiferromagnetic properties of MBT, the magnitude of $M_{n,\textrm{bulk}}$ highly depends on the parity of the layer number of MBT $N$.  $M_{n,\textrm{bulk}}$ is usually very small when $N$ is even [Fig.~\ref{supp-bulk-TSC}(f)]. This makes the corresponding topological region also very small, about 0.87 meV [Fig.~\ref{supp-bulk-TSC}(d)]. We don't plot this topological region in Fig.~3(e)(f) 
of the main text.
It is noted that the signs of the $C$ stem from surface states and bulk states are different when $N$ is even. This is because the gate-induced electrostatic potential will confine the bulk states closer to the top surface of MBT. Thus, the BSSs and bulk states have opposite magnetization (The sign of the Chern number is determined by the direction of magnetization). We also find that the magnetization of bulk states changes with the gate voltage because of the non-uniform distribution of the
electrostatic potential in MBT.  When the gate voltage is very negative, bulk states can also have large magnetization even for even $N$. As shown in the Fig.~\ref{fig-phase-diagram}(a), the induced superconducting gaps in topological regions stemming from bulk states are very small,
which is not favorable for achieving robust CMMs. 

\section{The electrostatic potential narrows the topological regions}\label{Appendix E}
\setcounter{equation}{0}
\renewcommand\theequation{E\arabic{equation}}
For simplicity, we consider the 2D effective Hamiltonian of MBT which consists of the Dirac-type surface states only. It takes the form 
$\mathcal{H}(\boldsymbol{k}) = \sum_{\boldsymbol{k}}\psi_{\boldsymbol{k}}^{\dagger}H_{sf}(\boldsymbol{k})\psi_{\boldsymbol{k}}$ with 
\begin{eqnarray}
    H_{sf}(\boldsymbol{k}) &=& v_F k_y\sigma_zs_x -v_F k_x\sigma_zs_y+m_k\sigma_0s_z \\ \nonumber
    &+& M\sigma_zs_z+V\sigma_zs_0,
\end{eqnarray}
where the field operator $\psi_{\boldsymbol{k}} = (c_{t\uparrow},c_{t\downarrow},c_{b\uparrow},c_{b\downarrow})^{T}$, $t$ and $b$ denote the TSSs and BSSs. $\uparrow$ and $\downarrow$ represent spin-up and spin-down, respectively. $\boldsymbol{k} = (k_x,k_y)$. $v_F$ is the Fermi velocity of surface states. $\sigma_i$ and $s_i~(i=x,y,z)$ are the Pauli matrix acting on layer and spin space, respectively. $m_k = m_0 +m_1(k_x^2+k_y^2)$ describes
the tunneling effect between TSSs and BSSs and set $m_1>0$. 
$M$ is the magnetization of surface states. Here layer number of MBT is odd (For even layers, the exchange field term changes as $M\sigma_0s_z$). $V$ is structure inversion asymmetry imposed by the gated induced electrostatic potential~\cite{WangJi2016}. 

The Bogoliubov-de Gennes (BdG) Hamiltonian for
the s-wave superconductor proximity coupled MBT is $\mathcal{H}_{\textrm{BdG}}(\boldsymbol{k}) =\sum_{\boldsymbol{k}}\Psi_{\boldsymbol{k}}^{\dagger}H_{\textrm{BdG}}(\boldsymbol{k})\Psi_{\boldsymbol{k}}/2$, with $\Psi_{\boldsymbol{k}} = (\psi_{\boldsymbol{k}},\psi_{\boldsymbol{-k}}^{\dagger})^{T}$ and
\begin{equation}\label{eq-surface}
H_{\textrm{BdG}}(\boldsymbol{k}) = 
\begin{pmatrix}
H_{sf}(\boldsymbol{k})-\mu & \Delta(\boldsymbol{k})  \\
\Delta^{\dagger}(\boldsymbol{k}) & -H_{sf}(-\boldsymbol{k})^{*}+\mu
\end{pmatrix},
\end{equation}
where $\mu$ is chemical potential, $\Delta(\boldsymbol{k})$ is the pairing function given by
\begin{equation}
\Delta(\boldsymbol{k}) = 
\begin{pmatrix}
i\Delta_{t}s_y &  0 \\
0 & i\Delta_{b}s_y
\end{pmatrix},
\end{equation}
where  $\Delta_{t(b)}$ is the SC gap of TSSs (BSSs).

We consider a limiting cases with $m_k = 0$. The Hamiltonian $H_{\textrm{BdG}}(\boldsymbol{k})$ is decoupled into two parts which contains BSSs and TSSs, respectively. The Chern number of these two surface states $\mathcal{C}_{t(b)}$ is determined by
\begin{equation}
\mathcal{C}_{t(b)} = 
\begin{cases}
\textrm{Sign}(M) & M^2 > \Delta_{t(b)}^2 + (\mu \mp V) ^2 \\
0 &  \textrm{Otherwise}
\end{cases},
\end{equation}
 The sign of $\mathcal{C}_{t(b)}$ is opposite (the same) in even (odd) layers of MBT. In experiments, we usually have $V \gg M,~\Delta_{t(b)}$~\cite{Zhang-NP-2010}. This makes the topological region stemming TSSs and BSSs well separated. As discussed in the main text, the Fermi level of TSSs can not be well tuned because of the band bending effect. This makes the magnetic gap of TSSs always below the Fermi level during the gate tuning. Thus, we always have $\mathcal{C}_{t} = 0$, which highly narrows  the topological regions, as show in Fig.~\ref{fig-phase-diagram}(c).

\bibliographystyle{apsrev4-1}
\bibliography{ref}

\end{document}